\documentstyle[psfig]{l-aa}
\begin{document}

\thesaurus {06 (08.09.2 (U Scorpii), 08.14.2)}

\title {\it Letter to the Editor \\ \smallskip \bf
The 1999 outburst of the eclipsing and recurrent nova U Scorpii\thanks{Based 
on observations collected with the telescopes of the European
Southern Observatory at La Silla, Chile and of the Padova \& Asiago
Astronomical Observatories (Italy)}}

\author {U.Munari\inst{1,3} \and T.Zwitter\inst{2} \and T.Tomov\inst{1,3} 
\and P.Bonifacio\inst{4} \and P.Molaro\inst{4} \and P.Selvelli\inst{5}  
\and L.Tomasella\inst{1} \and A.Niedzielski\inst{6} \and A.Pearce\inst{7} }
\offprints{U.~Munari (munari@pd.astro.it)}
\institute{
Osservatorio Astronomico di Padova, Sede di Asiago, 
I-36032 Asiago (VI), Italy 
\and
University of Ljubljana, Department of Physics, 
Jadranska 19, 1000 Ljubljana, 
Slovenia 
\and
CISAS - Center of Studies and Activities for Space, Univ. of Padova
``G.Colombo", Italy
\and
Osservatorio Astronomico di Trieste, Via Tiepolo 11, I 34131 Trieste, Italy
\and
CNR-GNA-Osservatorio Astronomico di Trieste Via Tiepolo 11, I 34131 Trieste, Italy
\and
Centre for Astronomy, N.Copernicus University, ul. Gagarina 11, PL-87100
Torun, Poland
\and 32 Monash Ave, Nedlands, WA 6009, Australia
}

\date{Received date..............; accepted date................}

\maketitle
\markboth{Munari et al.: The 1999 outburst of the recurrent nova 
  U Sco}{Munari et al.: The 1999 outburst of the recurrent nova U Sco} 
\begin{abstract}

The spectroscopic and photometric evolution of the 1999 outburst of the
eclipsing and recurrent nova U~Sco is presented. The photometric evolution
closely matches that of the previous events. The FWZI=10,000 km sec$^{-1}$
for emission lines at maximum has decreased to 4000 km sec$^{-1}$
by day +23, with continuous and dramatic changes in the line profiles. No
nebular line has become visible and the ionization degree has increased 
during the brightness decline. A not previously reported and quite
puzzling splitting of the emission lines into three components after the
first two weeks is outstanding in our spectra. The radiated luminosity is
found to be a tiny fraction of that of classical novae for any reasonable
distance to U~Sco.

\end{abstract}

\section{Introduction}

Previous outbursts of U~Sco were recorded in 1863, 1906, 1936, 1979 and
1987 (cf. Sekiguchi et al. 1988, hereafter S88). They were all characterized
by a very fast evolution ($t_3\sim 5$ days) and a large amplitude (from $V
\sim 18-19$ mag in quiescence to $V \sim 8$ mag at maximum). Several others
have been quite possibly missed because U~Sco lies just 4$^\circ$ from the
ecliptic. In the following the comparison will be limited to the 1979 and
1987 events, because these are the only ones for which useful spectroscopic
and photometric data have been obtained (S88, Barlow et al. 1981,
Williams et al. 1981 and Warner 1995; hereafter B81, W81 and W95 respectively).

The system in quiescence shows eclipses (Schaefer \& Ringwald 1995,
hereafter SR95) with a period of 1.2305631 days. The bright prospects to
derive the masses of the components from spectroscopic orbits have been
however hampered by the faintness in quiescence and devoted attempts with
4-m class telescopes have so far provided only contradicting results
(Johnston \& Kulkarni 1992, Duerbeck et al. 1993, SR95).

Thanks to an immediate notification by the outburst discoverer (P.Schmeer,
Belgium) and the VSNET network (cf.
http://www.kusastro.kyoto-u.ac.jp/vsnet/), we were able to begin the
monitoring of the 1999 outburst within a few hours from the maximum.  In
this letter we present the results of our all-out spectroscopic campaign and
report about the photometric evolution as long as U Sco has remained
brighter than V=15 mag.  Modeling of the data and additional observations of
U Sco once it will have returned to flat quiescence will be presented
elsewhere (Selvelli et al. 1999, in preparation). Similarly, a detailed
discussion of the reddening as inferred from the interstellar absorption
lines visible in our Echelle spectra will be addressed in detail elsewhere
(Munari and Zwitter 1999, in preparation).

\begin{table}
\caption[]{Journal of the spectroscopic observations. $\bigtriangleup t$ =
days since maximum brightness (see text); JD$_\odot$ = JD$_\odot$ -- 2451200; $\Phi$
= orbital phase from SR95; {\sl Tel.} = telescope code ($a$ = Asiago 1.22m +
B\&C +CCD; $b$ = Asiago 1.82m + Echelle +CCD; $c$ = ESO NTT + EMMI); {\sl
disp.} = dispersion (in \AA/pixel) around H$\alpha$; {\sl FWZI}= full width at zero intensity
for the H$\alpha$ profile (H$\beta$ for the first spectrum).}
\centerline{\psfig{file=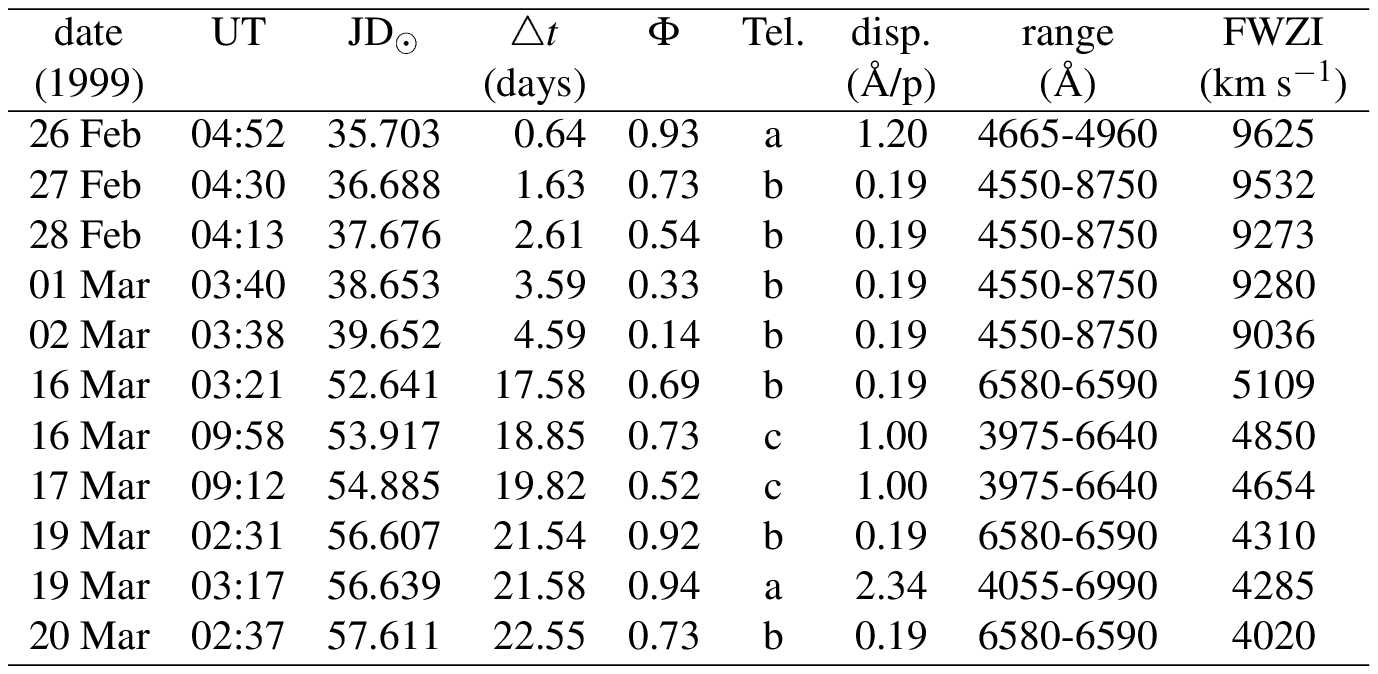,width=8.8cm}}
\end{table}

\section{Observations}

A journal of the spectroscopic observations is given in Table~1. In this
paper the dates are counted from the outburst maximum brightness $V~=~7.6$ mag
occurred on JD=2451235.062 (cf. IAU Circ. 7113).

High resolution spectra have been obtained with the Echelle spectrograph
mounted at the Cassegrain focus of the 1.82 m telescope which is operated by
Astronomical Observatory of Padova on top of Mt. Ekar, Asiago (Italy). The
detector was a Thomson THX31156 CCD with 1024$\times$1024 pixels,
19$\mu$m each, and the slit width was set to $1''.5$. The multi-order
Echelle spectra have been successfully processed into single dispersion ones
by using spectra of some unreddened A0~V stars close to U~Sco and observed
under identical conditions (inter-comparison of the similarly
processed A0~V star spectra shows the rectification and joining technique to
be accurate to better than $\sim$3\%). The gaps in the red/near-IR visible
in the spectra of Figure~2 are caused by non overlapping Echelle orders.

\begin{table}[!t]
\caption[]{Visual magnitudes of U Sco during the 1999 outburst. 
JD$_\odot$ = JD$_\odot$ -- 2451200; $\dag$ =
pre-outburst observations from VSNET archives; $\ast$ = data from IAU Circ.
7113.}
\centerline{\psfig{file=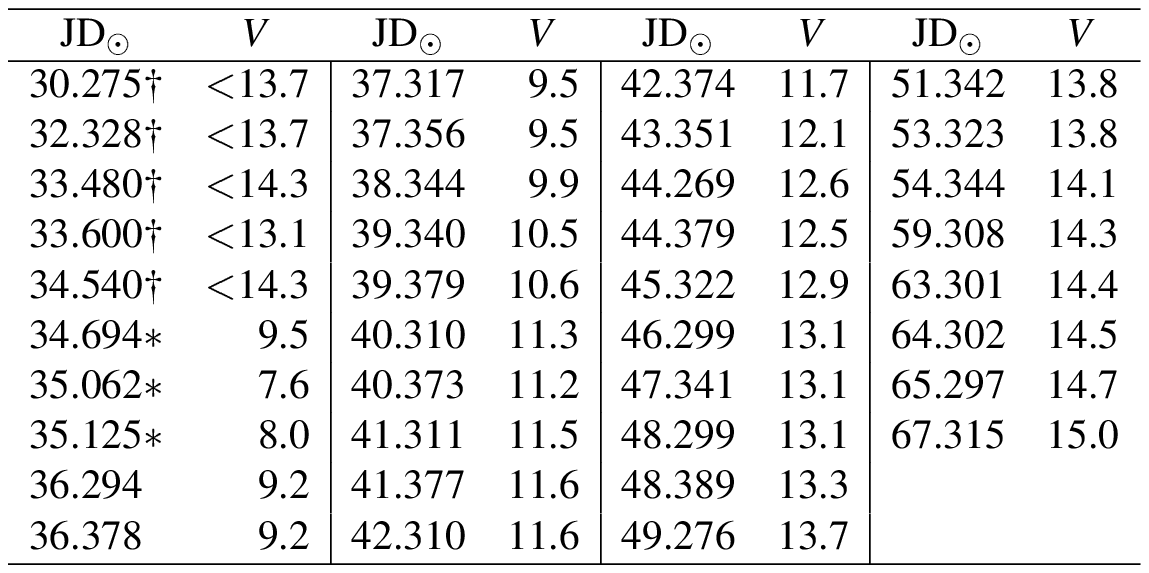,width=8.8cm}}
\end{table}
\begin{figure}
\centerline{\psfig{file=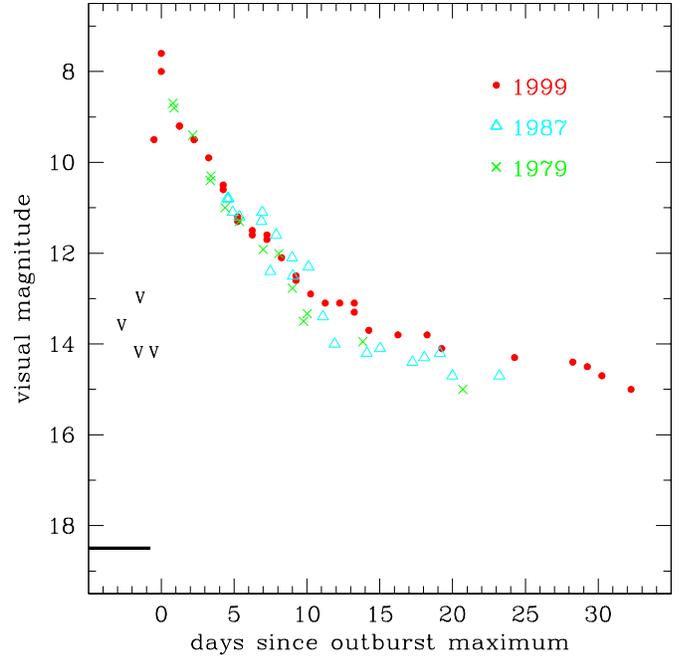,width=8.8cm}}
\caption[]{Comparison of the light curves of the last three outbursts
of U Sco (data from Table~2, B81 and S88). The thick line is the
average brightness in quiescence. Negative detections for 1999 
are given as {\sl fainter-than} symbols.}
\end{figure}

Medium dispersion spectra were secured with the B\&C+CCD spectrograph
at the 1.22 m telescope of the Astronomy Dept., Univ. of Padova, located in
Asiago too. The detector was a Wright Instr. CCD camera with a 512x512
pixel, 23$\mu$m size, UV-coated chip. The slit width was set to $2''$.

On March 16th and 17th U Sco was observed with the ESO NTT and the EMMI
instrument. The dispersing element was grism \#5 on the 16th, and
grisms \#5 and \#2 the following night. A $1''$ slit was used on both
nights.  The seeing was $ 0''.9$ the first night and $ 0''.8$ the
second. The DA white dwarf HIP~80300 was observed as flux standard on
March~17.

Visual estimates of U~Sco brightness have been obtained by one of us (AP)
with a private 16" f/4 reflector using the A.Henden (USNO) comparison
sequence as distributed by VSNET.  The visual estimates are listed in
Table~2 and the photometric evolution of the 1999 outburst is compared with
those of 1979 and 1987 events in Figure~1.

\section{Photometric evolution}

The photometric evolution of the 1999 outburst has followed quite closely
that of previous events as Figure~1 clearly shows. Small differences may be
easily accounted for by ($a$) different comparison sequences, and ($b$) a
mixture of different observing techniques used by B81 and S88 (visual
estimates directly at the eyepiece or on the screen of TV telescope guiding
systems, photography, etc.)

\begin{figure*}[!t]
\centerline{\psfig{file=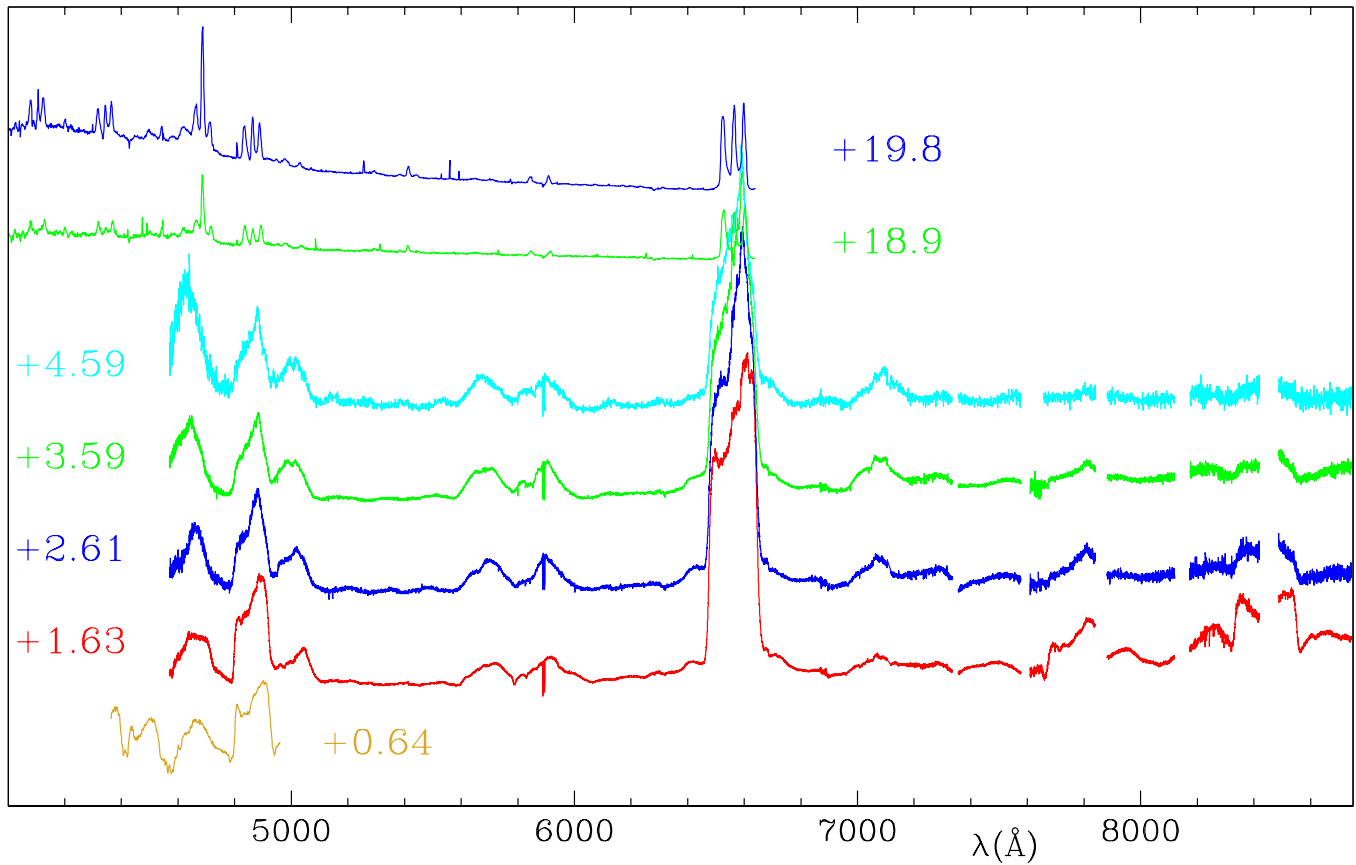,width=18cm,angle=270}}
\caption[]{Spectroscopic evolution of the 1999 outburst of U Sco from some
of our spectra. Numbers are days since maximum brightness. The spectra are
normalized to 1.0 at 5500 \AA\ and shifted to avoid overplotting. H$\alpha$
profiles are plotted on an expanded scale in Figure~3.}
\end{figure*}
\begin{figure}[!h]
\centerline{\psfig{file=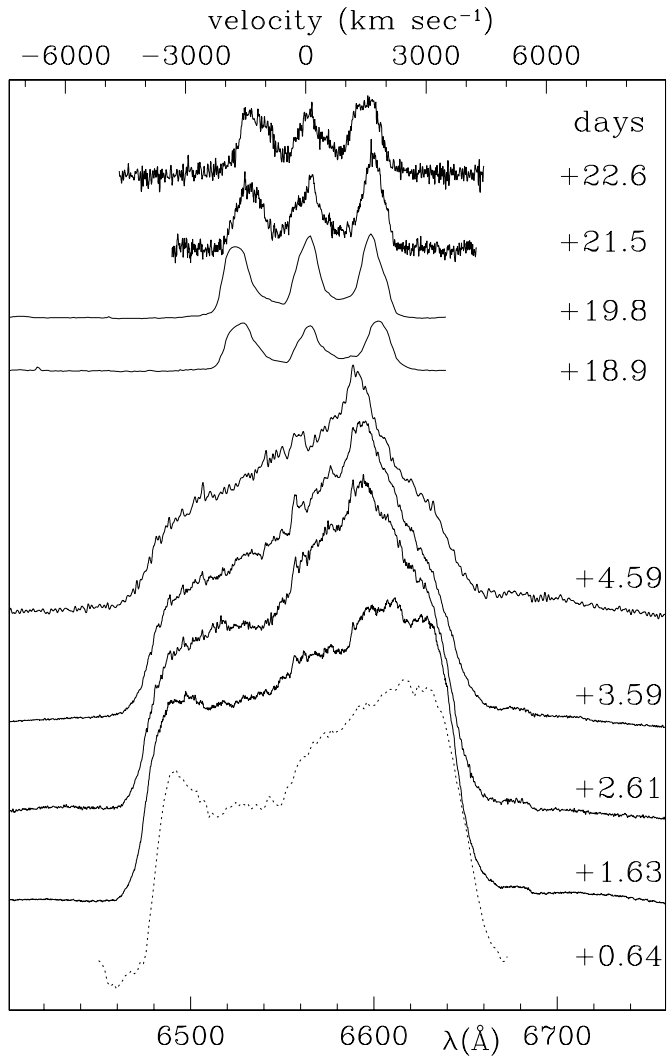,width=8.8cm}}
\caption[]{Evolution of the H$\alpha$ profile. The first spectrum (dotted
line) is actually a H$\beta$ profile (cf. Figure~1). The transition from a
saddle-like profile at earliest stages toward a more Gaussian-like one and
eventually to a triple-peaked shape is evident as it is the continuous
decrease in width (cf. Figure~4).}
\end{figure}
\begin{figure}[!h]
\centerline{\psfig{file=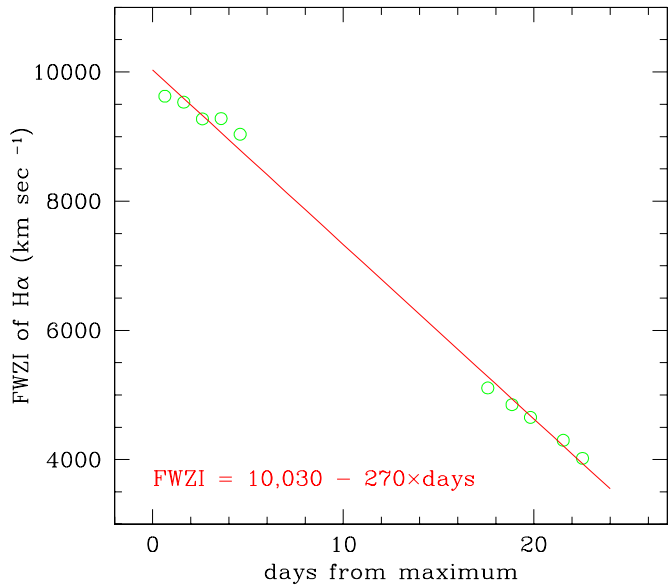,width=8.8cm}}
\caption[]{Decrease of the FWZI of H$\alpha$ since outburst maximum.}
\end{figure}

With a maximum brightness of $V=7.6$ on February 25.562, the 1999 outburst
has been characterized by a fast decline with $t_2 = 2.2$ and $t_3=4.3$ days,
very close to the 0.67 mag day$^{-1}$ reported by Payne-Gaposchkin (1957)
for the 1863 and 1936 events. The outburst was discovered by P.Schmeer on
February 25.194 when he estimated U~Sco at $V=9.5$. This suggests a fast
rise to maximum of the order of $\bigtriangleup$mag = 5.2 day$^{-1}$, or even
faster if the observation at $V=7.6$ on February 25.562 was actually past
the true maximum.

There is an important negative detection on Feb 25.040 listed in Table~2,
when U~Sco was found fainter than V=14.3. This is just 22 minutes before
central eclipse according to the SR95 ephemeris and 3.7 hours earlier than
the $V=9.5$ outburst discovery by P.Schmeer. Adopting the
$\bigtriangleup$mag = 5.2 day$^{-1}$ rise rate just estimated, U~Sco should
have been at $V=10.3$ mag at the time of the Feb 25.040 negative detection.
This seems to suggest that the dimensions of the outbursting WD were still
smaller than those of the occulting companion 0.522 days before maximum.
However, a different explanation is in order if the SR95 ephemeris should
turn out to be no more accurate in 1999 and/or the predicted minima are the
eclipse of the hot spot and not those of the WD.

Integrating the lightcurve in Figure~1, the energy radiated in the V band by
U Sco during the time covered by the observations in Table~2 can be
expressed as
\begin{equation}
E^{\rm rad}_{\rm V} = 2.65 \times 10^{40}\ D^2 \ 10^{0.4 \times A_V}  ~~~~~~~~{\rm ergs} 
\end{equation}
where $A_V$ is the extinction in magnitudes and $D$ is the distance in kpc.
The slope of the continuum in our spectra suggests a color temperature
of about 2$\times 10^4$ $^\circ$K. Assuming for sake of 
discussion that U~Sco has radiated on the average as a Kurucz's model atmosphere
with T=20,000 $^\circ$K  and log g = 3.0, we find that the
global radiated energy is $\sim 20 \times$ that radiated in the $V$ band,
so Eq.(1) can be rewritten for the bolometric energy as
\begin{equation}
E^{\rm rad} \sim 5 \times 10^{41}\ D^2 \ 10^{0.4 \times A_V}  ~~~~~~~~{\rm ergs} %
\end{equation}
For any reasonable distance inside the Galaxy and the extinction generally
adopted ($A_V$=0.6, cf. S88), the 1999 outburst of U~Sco appears
considerably underluminous compared to those of classical novae (cf. W95).
The same conclusion was reached by W81 from IUE observations of the 1979
outburst of U~Sco.

\section{Spectral evolution}

In comparison to previous outbursts our observations begun much earlier,
significantly extended toward later phases and have been performed at a
higher resolution and over a broader wavelength range (see Figure~2).

As for previous outbursts (cf. B81, S88), the spectrum has been
characterized by very wide emission lines, with Balmer hydrogen lines being
the strongest at earliest phases (in quiescence hydrogen lines are generally
absent and mimicked by the Pickering series of HeII, cf. Hanes 1985 and
Johnston \& Kulkarni 1992).

A feature not reported for previous outbursts (perhaps due to the poorer
resolution and looser time coverage) is the monotonic decrease with time of
the FWZI (full width zero intensity) of H$\alpha$ (and similarly for the
other Balmer lines; cf. Figures~2 and 3). The values in Table~1 are plotted
in Figure~4 where the linear fit is given by the equation:
\begin{equation}
FWZI (H_\alpha) \ = \ \ 10,030 \ - \ 270 \times {\rm days}   
                    \ \ \ \ \ \ \ \ \ \ \ \  {\rm km \ sec^{-1}}
\end{equation}
This linear decrease is difficult to explain in term of ejecta deceleration
by circumstellar material because the spectra carry no sign of the typical
signatures that characterize the presence of shock fronts (cf. Osterbrock
1989).

The identification of emission lines other than the Balmer ones is
complicated by the their large width. Surely present close to maximum are OI
7775 and 8446 \AA, HeI 5876 and 7075 \AA\ (the presumably weaker 6678 \AA\
line is nearly lost in the H$\alpha$ wings) and NII 5675. The complex at
5015 \AA\ could be a blend of HeI and NII, and NIII and CIII are probably
the main contributors to the blend at $\sim$4630 \AA. The ionization degree
has increased during the decline, with HeII 4686 \AA\ becoming visible after
the first week from maximum. As for previous outbursts, also this time the
nebular lines have not shown up in the late spectra of U~Sco, reinforcing
the notion that a limited amount of material -- if any -- has been ejected
by U~Sco.

The Balmer and O~I 7775-8446 \AA\ emission lines showed a saddle-like
profile at earliest phases, while other lines presented a more Gaussian-like
profile. At {\sl day +3} Balmer and O~I lines turned to single-peaked
profiles as well. Later evolution has been characterized by Balmer and HeII
lines to split into three components with velocity separation of the order
of $\pm$1600 km sec$^{-1}$. This triple--peak profile was not observed in
the 1987 outburst (cf. S88) and can be perhaps only marginally spotted in
the latest H$\beta$ profile presented by B81 for the 1979 outburst. The
eclipsing nature of U~Sco prevents an explanation of the triple peaks at
later phases (when the ejecta become presumably optically thin) as
collimated beams of material ejected at a large angle from the plane of the
orbit or of an accretion disc.

The 1979, 1987 and 1999 outbursts spectroscopically resemble each other only
in broad terms, with significant differences from eruption to eruption. 
Such differences might trace large changes from event to event in the
optical depth and kinematics of the ejecta. The optical depth must be
connected to the amount of ejected material (the velocity and time extent
remaining about the same from outburst to outburst), which should in turn
depend on the amount of material accreted between successive outbursts.

\end{document}